\documentclass[apj]{emulateapj}

\usepackage{natbib}
\usepackage{bm}		
\usepackage{mathptmx}	

\newcommand\arcsec{\mbox{$^{\prime\prime}$}}

\newcommand{\Sigc}{\Sigma_{\rm crit}}
\newcommand{\zsrc}{z_{\rm src}}
\newcommand{\kgrp}{{\kappa_{\rm grp}}}
\newcommand{\ggrp}{{\gamma_{\rm grp}}}

\begin{document}

\author{Neal Dalal\altaffilmark{1}} 
\affil{Institute for Advanced Study, Einstein Drive, Princeton NJ 08540}
\author{Casey R.\ Watson}
\affil{Dept.\ of Physics, The Ohio State University, Columbus OH 43210}

\altaffiltext{1}{Hubble Fellow}

\title{What are the environments of lens galaxies?}

\begin{abstract}
Using measured tangential shear profiles and number counts of massive
elliptical galaxies, the halo occupation distribution of strong lensing
galaxies is constrained.  The resulting HOD is then used to populate
an N-body simulation with lens galaxies, in order to assess the
importance of environment for strong lensing systems.  Typical
estimated values for the convergence and shear produced by nearby
correlated matter are $\kgrp\simeq\ggrp\simeq0.03$, with much stronger
events occurring relatively infrequently.  This implies that estimates
of quantities like the Hubble constant are not expected to be
significantly biased by environmental effects.  One puzzle is that
predicted values for the external shear at lens galaxies are far below
the values obtained by modeling of strong lensing data.
\end{abstract}


\maketitle

\section{Introduction}

Strong gravitational lensing has emerged as an important tool in
the astrophysical toolbelt.  Strong lenses have been used, for example, 
to measure
the Hubble constant \citep[e.g][and references therein]{schechter04}, 
to study the properties of elliptical galaxies
\citep[e.g.][]{rusin03}, to study dark matter substructure
\citep[e.g.][]{kd04}, to resolve the inner structure of quasars
\citep{BEL1004}, even to constrain the geometry of the universe
\citep{a1689}.  Many of the applications of strong lensing require
reconstructing the gravitational potential of the lens, which includes
contributions both from the principal lens galaxy and from any
(projected) nearby structures.  Models of strong lenses
often treat the lens galaxy itself in exquisite detail
\citep[e.g][]{cohn01}, however the lens environment is usually treated
more simply.  Except in cases with obvious nearby galaxies
\citep[e.g.][]{koopmans03,kundic97}, the environments of lenses are
typically modeled simply as tidal shear.

How valid is this approximation?  In some instances, higher order
terms in the potential than the $\sim r^2 \cos 2\theta$ term
represented by external shear may be required, and
\citet{koch_saasfee} describes quantitatively how to estimate the
importance of such terms for a lens model.  Additionally, the nearby
environment can contribute not only tidal shear, but (projected) mass
density as well. As discussed by \citet{holder03},
massive ellipticals are biased tracers of the
underlying mass density; we expect massive ellipticals to reside in
overdense environments, and indeed, many lens galaxies are found in
poor groups or galaxy clusters. \citet{keetonz04}, among others, 
point out that the environmental mass density, here
denoted $\kgrp$, can lead to a mass sheet degeneracy in many of the
quantities derived from strong lensing analyses.  For instance, the
Hubble constant inferred from time delays scales as
$H_0\propto(1-\kgrp)$.  In the age of precision cosmology, it is
clearly important to determine the magnitude of this uncertainty in
lens models.

In principle, our theory of structure formation should already provide
an answer to this question.  In the cold dark matter (CDM) model, the
clustering of mass proceeds largely through the growth and merging of 
dark matter halos.  Assessing the importance of lens environments
reduces basically to determining, where (in which halos) do lens
galaxies live?  Placing lens galaxies chiefly in the most massive,
biased objects like galaxy clusters will of course enhance the effects
of environment, whereas environment will be less important if lenses
live in lower mass halos.

We do not have complete freedom to pick the halo occupation distribution
of lens galaxies, however: there are now several observational
constraints on models for elliptical galaxies and their hosts.  First,
galaxy-galaxy lensing \citep{sheldon04,sdssbias} provides direct
constraints on the average mass profiles surrounding lens galaxies.
Second, choosing a model for the halo occupation distribution of galaxies
implicitly makes a prediction for the number density of those
galaxies, and there are now precise measurements of the number density
of massive elliptical galaxies (which most strong lenses are).
Accordingly, we can use these measurements to constrain the HOD of
lens galaxies, and thereby compute the effects of lens environments.

The plan of this paper is as follows.  In \S\ref{halo}, the number
counts of \citet{sheth03} and the galaxy-galaxy lensing results of
\citet{sheldon04} are combined, in the context of the halo model, to
constrain the halo occupation distribution of lens galaxies.  In
\S\ref{simul}, this HOD is used to populate an N-body simulation with
lens galaxies, which is then exploited to make predictions for the
effects of lens environments.

\section{Halo model calculation}
\label{halo}

\citet{hujain04} explain in detail how the halo model may be used to
compute the number density and average surface density (and shear) profiles
around galaxies.  As they discuss, the galaxy-convergence angular
correlation function may be computed by projecting the 3-D galaxy-mass
power spectrum using the \citet{limber} approximation,
\begin{eqnarray}
\langle\kappa\rangle(\theta)&=&\int\frac{l dl}{2\pi}J_0(l\theta)
C_l^{g\kappa} ,\\
C_l^{g\kappa}&=& \int_0^{\zsrc} dz W_g(z) W_\kappa(z) 
\frac{P_{g\delta}(k=l/r(z))}{r^2(z) c/H(z)}, \\
W_g(z)&=& {\bar n}_A^{-1}\;r^2(z)\frac{c}{H(z)} {\bar n}_V(z), \\
W_\kappa(z)&=& \frac{3}{2}\Omega_m H_0^{\;2}\frac{(1+z)r(z)}{c\,H(z)}
\frac{D_{LS}}{D_S}.
\end{eqnarray}
Here, ${\bar n}_V(z)$ is the volume number density of lens galaxies as a
function of redshift, 
${\bar n}_A=\int dz r^2(z)\frac{c}{H(z)} {\bar  n}_V(z)$ 
is the projected angular number density of lens galaxies, $r(z)$ is
the comoving angular diameter distance, and $D_{LS,S}$ are the angular
diameter distances from lens to source and from observer to source,
respectively.  

The halo model enters by providing a prescription for the galaxy-mass
power spectrum $P_{g\delta}$.  The ingredients are models for the dark
matter halo mass function $dn_h/dM$, the halo bias $b(M)$, the halo density
profile $\rho(r)$, and the halo occupation distribution of galaxies.
For DM halos, we use the mass function and bias prescription of
\citet{shethtormen}, the density profile of \citet{nfw}, and halo
concentrations of \citet{bullock01}.  The halo occupation distribution
(HOD) specifies how galaxies populate DM halos.  \citet{kravtsov04}
present a model for the HOD based upon N-body simulations.  In their
model, the mean number of galaxies (say, brighter than luminosity $L$)
residing in halos of mass $M$ has two pieces, 
${\bar N}(M)={\bar N}_c(M)+{\bar N}_s(M)$,
corresponding to central galaxies and satellite galaxies.  They find
that $N_c(M)$ is well fit by a step function, 
$N_c(M)=\Theta(M-M_{\rm th})$ where the parameter $M_{\rm th}$ is the
minimum mass halo capable of hosting a central galaxy with the
required properties (e.g.\ luminosity).  \citeauthor{kravtsov04}\ also
find that the satellite term is well described by a Poisson distribution
with mean 
\begin{equation}
{\bar N}_s(M) = \left(\frac{M}{A_s M_{\rm th}}\right)^{m_s}
\label{nsat}
\end{equation}
for $M>M_{\rm th}$ and zero otherwise.
The halo mass function and HOD together specify the predicted galaxy
volume number density, viz
\begin{equation}
{\bar n}_V(z)=\int dm \frac{dn_h}{dM}\,{\bar N}(M).
\label{num}
\end{equation}
Similarly, the halo model expression for the galaxy-mass power
spectrum becomes, in the notation of \citet{hujain04}, 
\begin{equation}
P_{g\delta}(k)=I_{1g}(k)I_{1m}(k)P_{\rm lin}(k) + I_{2c}(k),
\label{pgd}
\end{equation}
where 
\begin{eqnarray}
I_{1m}(k) &=& \int dM \frac{M}{\rho_M}\,\frac{dn_h}{dM}\,b(M) y_h \\
I_{1g}(k) &=& {\bar n}_V^{\,-1} \int dM [{\bar N}_c+{\bar N}_s y_g]
\frac{dn_h}{dM} b(M) \\
I_{2c}(k) &=& {\bar n}_V^{\,-1} \int dM \frac{M}{\rho_M}\,
\left[\frac{dn_h}{dM}\,({\bar N}_c y_h+{\bar N}_s y_h y_g)\right.
\nonumber \\
&& \left.\qquad + \frac{dn_s}{dM} \Theta(M-M_{\rm th}) y_s\right],
\label{i2c}
\end{eqnarray}
$y_h$ is the normalized Fourier transform of the DM density profile, 
\begin{equation}
y_h(k,M) = M^{-1} \int_0^{r_{\rm vir}} dr\; 4\pi r^2 \rho(r)
\frac{\sin(k\,r)}{k\,r} ,
\end{equation}
and similarly $y_g$ is the normalized Fourier transform of the
satellite galaxy distribution within halos, here assumed to track the
mass, so that $y_g=y_h$.  Note that the
redshift-dependent concentration is implicitly included in the density
profile $\rho(r,M)$.  The satellite mass function is given by
\begin{equation}
\frac{dn_s}{dM}=\int_M^{\infty} dM_h\,m_s 
\left(\frac{M_h}{A_s M}\right)^{m_s} \frac{dn_h}{d M_h}.
\end{equation}

The observables to match are the number density and tangential shear
profile.  Eqn.~\ref{num} gives the predicted number density, while the
predicted shear profile is given by
\begin{equation}
\langle\gamma_T\rangle(\theta) = \int \frac{l\,dl}{2\pi}
C_l^{g\kappa} J_2(l\theta).
\end{equation}
Strong gravitational lenses tend to be massive elliptical galaxies,
and \citet{sheldon04} report the average shear profile of elliptical
galaxies with velocity dispersion $\sigma>182$ km/s observed in the
Sloan Digital Sky Survey (SDSS).  \citet{sheth03} have measured the
velocity dispersion function of SDSS ellipticals, and found it to be
well fit by a Schechter-type function.  For ellipticals with
$\sigma>182$ km/s, their fit gives a total number density of 
$n_{\rm obs}=0.00172 ({\rm Mpc}/h)^{-3}$.  Also, the mean squared
velocity dispersion for this sample is
$\langle\sigma^2\rangle=(216\rm{km/s})^2$.  For a lens at $z=0.45$ and
source at $\zsrc=2$, this would correspond to an Einstein radius of 
$\theta_{\rm E}=4\pi (\sigma/c)^2 D_{LS}/D_S=0.9\arcsec$, in good
agreement with typical lens splittings.

\begin{figure}
\plotone{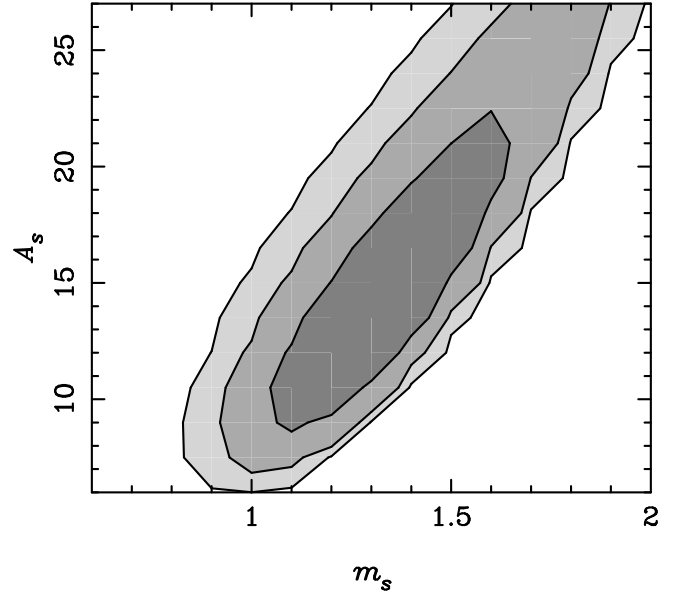}
\caption{Plotted are 68, 95, and 99\% likelihood contours in the
  $m_s-A_s$ plane for $M_{\rm th}=10^{12.5} h^{-1} M_\odot$.  
\label{hodpars}}
\end{figure}

The ingredients are now in place to find the HOD parameters which best
fit the observed data on massive ellipticals.  This HOD will be used
in the next section to populate halos found in an N-body simulation,
so it is important that consistent definitions of halo mass are used
for the halo model calculation and the group finder, the
friends-of-friends algorithm with linking length $b=0.2$.
\citet{jenkins01} have found that FOF with $b=0.2$ corresponds well to
halo overdensity of 180, and that the Sheth-Tormen mass function
provides a good fit to the mass function if the halo mass is defined
not as the virial mass, but as $M=M_{180}$, i.e.\ the mass enclosed
within radius $r_{180}$ for which the average interior density is 180
times the mean matter density $\rho_M$.  The \citet{bullock01} model
for halo concentrations is given in terms of the virial mass, not
$M_{180}$ which is typically $\sim1.2\times$ larger than 
$M_{\rm vir}$.  Given the NFW density profile, it would not be
difficult to translate from $M_{180}$ to $M_{\rm vir}$
\citep{hukravtsov}, but the dependence of concentration on mass is so
weak ($c_v\propto M^{-0.13}$) that this distinction is ignored here.

\begin{figure}
\plotone{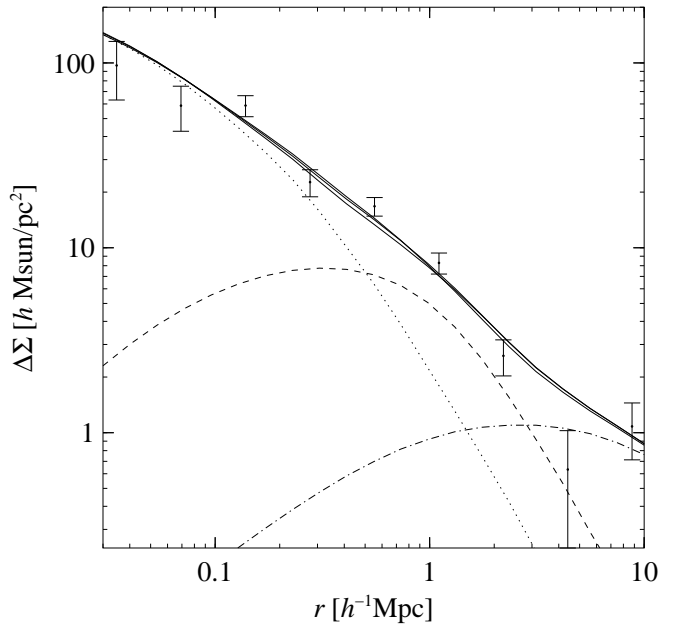}
\caption{The galaxy-shear correlation function for massive
  ellipticals.  Data points are from \citet{sheldon04}, kindly
  provided in electronic form by D.\ Johnston (2004, priv.\ comm.)
  The solid curves correspond to the predicted
  $\Delta\Sigma=\Sigc\langle\gamma_T\rangle$ for three different halo
  occupation distributions, with $A_s=15$, $m_s=1.4$; 
  $A_s=21$, $m_s=1.6$; and $A_s=10.5$, $m_s=1.2$.  The other curves
  show the contributions to eqns.~(\ref{pgd}) and (\ref{i2c}): the
  galaxies' own halos or subhalos (dotted), group halos (dashed), and
  two-halo term (dash-dot).
\label{avekap}}
\end{figure}

The HOD used here has three parameters: $M_{\rm th}$, $A_s$ and
$m_s$.  Figure~\ref{hodpars} shows likelihood contours in the
$A_s-m_s$ plane, for $M_{\rm th}=10^{12.5} h^{-1} M_\odot$, assuming a
10\% uncertainty for the number density $n_{\rm obs}$. Other
values of $M_{\rm th}$ gave significantly poorer fits, so $M_{\rm th}=
10^{12.5} h^{-1} M_\odot$ will be assumed throughout.  As can be seen
from the figure, a broad region in the $A_s-m_s$ plane, centered on
$A_s=15$, $m_s=1.4$ provides reasonable fits to the data.  These
values compare reasonably well to HOD parameters derived by
\citet{zheng04} from hydrodynamic simulations and semi-analytic models.
The satellite fraction for these HOD parameters is roughly 25\%, in
good agreement with the estimates of \citet{keeton00}.
There appears to be a mild degeneracy between $A_s$ and $m_s$, in the
sense that increasing $A_s$ (which lowers the fraction of lens
galaxies that are satellites) may be compensated by increasing $m_s$
(which places satellites in more massive groups) to hold fixed the
tangential shear profile; see fig.~\ref{avekap} for an example.  However
note that large $m_s$ is likely disfavored by measurements
of the multiplicity function in clusters \citep{kochanek03}.

One defect which has been glossed over so far is that the SDSS data on
elliptical galaxies used to constrain the HOD parameters extend to
redshifts $z\lesssim0.15$, much less than typical lens redshifts 
$z_l\sim0.5$.  In order to apply this HOD to the lens population, it
must be assumed that there is little evolution in massive ellipticals
between $z=0.15$ and $z=0.5$.

Note that the average $\kgrp$ may be calculated using these HOD
parameters using just the halo model, by keeping only the term in 
$P_{g\delta}$ corresponding to the correlation between satellite
galaxies and mass in their parent halos:
\begin{equation}
P_{g\delta}^{\rm grp}(k)=\frac{1}{{\bar n}_V}\int dM \frac{M}{\rho_M} 
\frac{dn_h}{dM}\,{\bar N}_s\,y_h\,y_g.
\label{kgroup}
\end{equation}
This somewhat underestimates $\kgrp$, because in this
formalism halo profiles extend only out to $r_{180}$, whereas in
reality there is mass overdensity extending well beyond this radius
\citep[e.g.][]{guzikseljak}.  With this caveat in mind, applying
eqn.~\ref{kgroup} gives an average $\kgrp\approx0.02$.  As noted
above, fractional errors in (e.g.) the Hubble constant will be 
${\cal O}(\kgrp)$, implying that estimates of the Hubble constant
which ignore the group convergence make $\sim 2\%$ errors for most
lens systems, which is still small compared to other systematics.

\section{Application to N-body simulations} \label{simul}

The halo occupation distribution from the previous section can now be
used to populate halos from an N-body simulation with lens galaxies,
to compute the distribution of lens environments.  A small
(by modern standards) dissipationless simulation with $256^3$
particles in a 256 $h^{-1}$Mpc box was run using the TPM 
code\footnote{http://www.astro.princeton.edu/$\sim$bode/TPM}
\citep{tpm} for a 
WMAP cosmology with $\Omega_M=0.27$, $\Omega_\Lambda=0.73$ and
$h=0.72$.  Initial conditions were generated using the GRAFIC1 
code\footnote{http://arcturus.mit.edu/grafic} \citep{grafic}, for a
scale-invariant primordial power spectrum ($n=1$) and present-day
amplitude $\sigma_8=0.9$.  The particle mass for this simulation was
$m_p=7.49\times 10^{10} h^{-1} M_\odot$, so that halos of mass
$M=10^{12.5} h^{-1} M_\odot$ contained 42 particles.  Since the
simulation was dissipationless, $\Omega_B=0$ was assumed for
generating the initial conditions. 

The $z=0.45$ output from this simulation was then populated with lens
galaxies using the best fitting HOD from the previous section, with
$M_{\rm th}=10^{12.5}h^{-1}M_\odot$, $A_s=13.5$, and $m_s=1.3$. First,
the friends-of-friends algorithm\footnote{The implementation of FOF
provided at the UW N-body shop was used,
http://www-hpcc.astro.washington.edu/tools/fof.html.}
with linking length $b=0.2$ was used to find halos in the simulation.
Central galaxies were placed at the most bound particles in these
halos.  The number of satellites for each halo was determined by
generating random numbers from a Poisson distribution with mean given
by eqn.~\ref{nsat}.  Satellite galaxies were assigned the positions of
randomly selected particles in each halo, to match the assumption
above that the satellite distribution follows the DM distribution.

\begin{figure*}
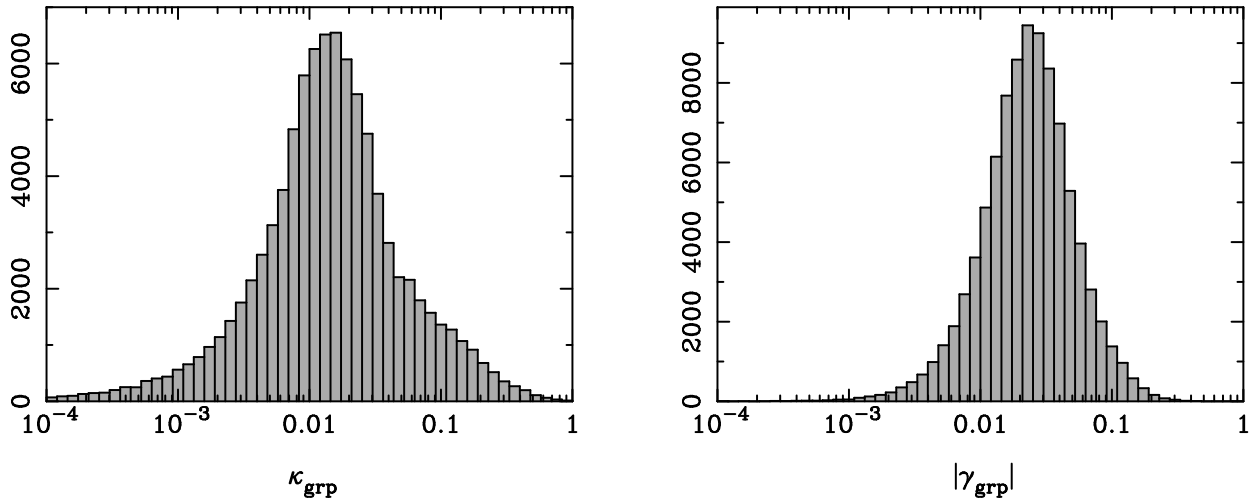

\centerline{\includegraphics[width=0.42\textwidth]{kap.eps}
\qquad \qquad\includegraphics[width=0.42\textwidth]{gam.eps}}
\caption{Histograms of (left) $\kgrp$ and (right) $|\ggrp|$ for the
  $z=0.45$ simulation, assuming $\zsrc=2$.  The HOD used here had
  $M_{\rm th}=10^{12.5} h^{-1} M_\odot$, $A_s=13.5$, and $m_s=1.3$.
\label{histo}}
\end{figure*}

Next, convergence and shear (assuming a source redshift $\zsrc=2$)
were computed at the galaxy positions 
by projecting the particle data along three orthogonal
axes.  For central galaxies, the contribution from the galaxies' own halos
must be removed in order to compute solely environmental effects.
This was accomplished by excluding, for each galaxy, all particles in
its halo (as defined by FOF) when computing the projected convergence.
Applying this method for shear would be computationally onerous,
since the shear is computed by repeated Fourier transforms of the full
convergence map.  Instead, to compute the environmental shear at each
central galaxy, comoving 100 $h^{-1}$kpc cavities were excised around each
galaxy.  This truncated particle list was then projected to make a
convergence map with comoving 31.25 $h^{-1}$ kpc resolution, which was
then Fourier transformed to produce shear maps of the same
resolution. The resulting shear values at the galaxy positions were
mildly dependent upon the 
precise cavity radius; using comoving 50 $h^{-1}$kpc cavities instead
of 100 $h^{-1}$kpc cavities gave shear values $\sim 20\%$ larger.  
For satellite galaxies, no such excision of particles was required,
because the satellites were placed randomly inside their parent
groups, not on subhalos. The resulting histogram of
$\kgrp$ and $\ggrp$ values is plotted in figure~\ref{histo}.

There are a few points to note.  First, the mean values of convergence
and shear are similar, with
$\langle\kgrp\rangle\approx\langle|\ggrp|\rangle\approx0.03$.  This
agrees roughly with the simple halo model estimate for
$\langle\kgrp\rangle$ given in the previous section.  While the
$\ggrp$ histogram appears well described by a lognormal distribution,
the $\kgrp$ histogram exhibits extended tails.

An interesting related question is that of the frequency of high $\kgrp$
or $\ggrp$.  In these simulations, $\kgrp>0.1$ occurred with roughly
$\sim6\%$ probability, while $|\ggrp|>0.1$ had $\sim3\%$
probability.  One difficulty in making these estimates is 
that the tails of the $\kgrp$ and $\ggrp$ distributions
are likely overrepresented, due to projections.  
Galaxies which project near unrelated massive groups
will appear to have higher $\kgrp$ and $\ggrp$ than is truly
representative for their own environments.  Similarly, galaxies which
project near voids will have overly low $\kgrp$.  
Of course, from the viewpoint of estimating errors on derived
quantities like the Hubble constant, such projections are just as
important to consider as the immediate environment of the lenses.

One last point to consider is that selection effects can modify the
probability distributions for $\kgrp$ and $\ggrp$ for actual lenses
compared to the distributions computed here.  For example, the strong
lensing cross section scales as $(1-\kgrp)^{-2}$, so in principle the
histograms of convergence and shear should be weighted by this
factor \citep{holder03}.  Reweighting the distributions 
(excluding the tiny fraction of galaxies with $\kgrp>0.5$), 
the average $\langle\kgrp\rangle$ increases to 0.037, while 
$\langle\ggrp\rangle$ increases to 0.033.  Another possible selection
effect is that ellipticals with higher velocity dispersion (and hence
larger lensing cross section $\propto\theta_{\rm E}^2\propto\sigma^4$)
will tend to live in more  massive, more biased halos, and so the
$\kappa$ and $\gamma$ histograms should be weighted by this factor as
well for lenses.  Unfortunately, this requires a more detailed halo
occupation distribution than has been used here.

\section{Discussion}

This paper has presented a calculation of the effects of the environments
surrounding lens galaxies.  Lens galaxies were placed in an
N-body simulation using a halo occupation distribution calibrated to
match both the number density and tangential shear profiles observed
for massive elliptical galaxies in the Sloan Digital Sky Survey.  The
SDSS results are measured for galaxies at low redshift,
$z\lesssim0.2$, while lens galaxies are typically at higher redshift,
$z\approx 0.5$, so the results presented here will be meaningful only
if there is little evolution in elliptical galaxies between $z\approx
0.15$ and $z\approx 0.45$.  The expected average values for
convergence and shear are 
$\langle\kgrp\rangle\approx\langle|\ggrp|\rangle\approx0.03$, with a
spread of $\sim0.6$ and $\sim0.35$ in $\log_{10}$ for $\kappa$ and
$\gamma$ respectively.  Values much higher than the mean appear
relatively rarely, with $\kgrp>0.1$ occurring with $\sim 6\%$
probability and $\ggrp>0.1$ occurring with $\sim3\%$ probability.  

Previous work has considered related topics.  Both \citet{kks97} and
\citet[HS]{holder03} have estimated the magnitude of tidal shear produced
by material correlated with lens galaxies.  This paper has followed an
approach quite similar to that developed by HS.
The results presented here appear similar to those of
\citeauthor{kks97}, however HS find somewhat
higher environmental shear.  This appears to be due to a different
HOD; HS use massive elliptical galaxies produced
using the semi-analytic recipes of \citet{gif}.  As they note, their
lens galaxies tend to live in very massive halos, with roughly
$\sim30\%$ residing in clusters more massive than $10^{14} h^{-1}
M_\odot$.  Accordingly, the number density of massive elliptical
galaxies predicted by HS is lower than the values computed here by
roughly an order of magnitude.
Further observations of the environments surrounding lens
galaxies \citep[e.g.][]{fassnacht04} can help determine whether lenses
typically do reside in such massive halos.

Recently, \citet{keetonz04} have investigated the importance of the
group convergence on strong lenses.  They construct a mock group including
several supercritical galaxies separated by a few Einstein radii.  By
explicit construction, they showed that in cases like these, $\kgrp$
can approach $\sim0.15-0.2$, which would dramatically skew the results
of analyses ignoring group convergence.  Clearly, for systems like
these, it is necessary to model such close neighbors using both
convergence and shear, and indeed in systems like PG 1115+080 the
nearby groups are modeled with isothermal spheres or NFW profiles
\citep[e.g.][]{treukoopmans02,kd04}.  For the majority of systems,
however, the environment is treated merely with tidal shear, and the
results of this paper indicate that such a
treatment is usually adequate.  If typical $\kgrp$ at lens galaxies is
$\sim3\%$, then errors in the Hubble constant will similarly be at
the $\sim3\%$ level, which is still unimportant compared to the
$\gtrsim10\%$ systematic errors typical of lensing $H_0$'s. 

One remaining puzzle from this work is the origin of the large shears
measured at lens galaxies.  As discussed by \citet{kks97}, successful
models of strong lenses typically require external shear at the level
of $\gamma\sim10-15\%$, considerably higher than the $\ggrp\approx
3\%$ found here. Such shear cannot be explained by uncorrelated
structure (e.g. line-of-sight projections), which should produce shear
of order the cosmic shear, $\sim1-2\%$ on these scales.  
HS have argued that such strong shear naturally
occurs for lenses living in massive systems like galaxy clusters where
$\kappa_s=\rho_s r_s/\Sigc\approx 0.15$.  On the other hand, 
\citeauthor{kks97}\ 
have suggested that much of this so-called external shear originates
in the halos of the lens galaxies themselves.  To support this idea,
they note that the total quadrupole at the images aligns quite well 
with the galaxy orientation, while misalignments larger than observed
would arise if $10-15\%$ external shear were completely uncorrelated
with the galaxy orientation \citep[see][for an updated version of
  this]{koch_saasfee}.  As noted by Holder 
(2003, priv.\ comm.), however, this argument has difficulty.  Halo
material stratified on ellipses with the same ellipticity as the mass
interior to the Einstein ring will produce no effect on the lens
images, as shown by Newton.  For the halo to produce this shear, it
must either have principal axes substantially twisted relative to the
galaxy principal axes, or it must have a very different ellipticity.
To illustrate, imagine taking an isothermal ellipsoid with
Einstein radius $r_{\rm E}$, and beyond some radius $r_0$, either
twisting the ellipsoidal axes by some angle $\Delta\theta$ or changing
the ellipticity by $\Delta\epsilon$.  Then the change in shear at the
Einstein radius, in the limit $r_0\gg r_{\rm E}$, will be of order 
\begin{equation}
\Delta\gamma\sim \frac{1}{2\pi}\frac{r_{\rm E}}{r_0}
\left(\Delta\epsilon + 2\epsilon\sin^2\Delta\theta\right),
\end{equation}
where typical lens Einstein radii are $\simeq 2 r_{\rm eff}$ and
typical ellipticities are $\epsilon\sim 0.3$.
\citet{hoekstra03} have found that galaxies and
their extended halos align quite well, with halo ellipticities on
average $\gtrsim0.7\times$ the observed galaxy ellipticity, which
limits $\Delta\gamma\lesssim 0.3 \epsilon r_{\rm E}/2\pi r_0$.
This casts doubt on a halo origin for observed external shear; 
$r_0\sim r_{\rm E}$ would be required to produce enough shear in this
way.  Therefore, dramatic changes in the density structure must occur
in lens galaxies just outside their Einstein radii, near the
transition between baryon-domination to DM-domination, for the lenses
themselves to account for the observed shear.  
On the other hand, halo substructure conceivably could account
for some of the unexplained shear \citep{cohn04}.

\acknowledgments{The authors thank Gil Holder, Dave Johnston, Chuck Keeton,
  Chris Kochanek, David Rusin and Zheng Zheng for many
  helpful discussions, and Paul Bode for help with the N-body
  simulations. ND is supported by NASA through Hubble
  Fellowship grant \#HST-HF-01148.01-A awarded by the Space Telescope
  Science Institute, which is operated by the Association of
  Universities for Research in Astronomy, Inc., for NASA, under
  contract NAS 5-26555.
}


\end{document}